\documentclass[reprint,pra,aps,nofootinbib,longbibliography,superscriptaddress]{revtex4-2}
\usepackage{amsmath,amssymb,amstext}
\usepackage[pdftex]{graphicx}
\usepackage{tabularx}
\usepackage{bm}
\usepackage{mathrsfs}
\usepackage{fancyhdr}
\usepackage{booktabs}
\usepackage{longtable}
\usepackage{multirow}
\usepackage{float}
\usepackage{outlines}
\usepackage{physics}
\usepackage{dsfont}

\usepackage{braket}

\usepackage{amsthm}
\usepackage{bbold}
\usepackage{comment}
\usepackage[dvipsnames]{xcolor}
\usepackage{quantikz}

\usepackage[textsize=footnotesize]{todonotes}
\usepackage[pdftex,pagebackref=false]{hyperref}

\hypersetup{
    unicode=false,          
    pdffitwindow=false,     
    pdfstartview={FitH},    
    pdfnewwindow=true,      
    colorlinks=true,        
    linkcolor=blue,         
    citecolor=green,        
    filecolor=magenta,      
    urlcolor=cyan           
}
\usepackage{cleveref}

\allowdisplaybreaks

\newtheorem*{proposition*}{Proposition}

\newcounter{todocounter}
\begin{document}
\title{Interpreting Quantum Learning Models via Stochastic Processes}
\author{Johannes Fankhauser}
\email{Johannes.Fankhauser@uibk.ac.at}
\affiliation{University of Innsbruck, Institute for Theoretical Physics, Technikerstrasse 21a, A-6020 Innsbruck, Austria}
\author{Lukas J. Fiderer}
\affiliation{University of Innsbruck, Institute for Theoretical Physics, Technikerstrasse 21a, A-6020 Innsbruck, Austria}
\author{Hans J. Briegel}
\affiliation{University of Innsbruck, Institute for Theoretical Physics, Technikerstrasse 21a, A-6020 Innsbruck, Austria}
\date{\today}     

\begin{abstract}

Quantum machine learning models define probabilistic input–-output maps through coherent quantum evolution and measurement. While such models can exhibit computational advantages, their internal functioning and decision making generally resists interpretation in terms of stochastic trajectories through intermediate configurations. In contrast to classical (Markovian) stochastic processes, quantum dynamics generically violates the Chapman–-Kolmogorov divisibility condition, preventing a decomposition into probabilistically meaningful intermediate transitions. We develop a probabilistic framework for representing quantum learning models as stochastic processes over configuration spaces where the dynamics are modeled as linear maps on probability distributions. Starting from a fixed POVM, arbitrary quantum channels induce transition kernels on the associated probability representation. For informationally complete POVMs, and in particular SIC-POVMs, these kernels are Markovian but generally quasi-stochastic, with non-classicality appearing as negativity. By contrast, projective spaces admit positive stochastic kernels but generally require non-Markovian dynamics due to the failure of Chapman---Kolmogorov divisibility. This yields a trade-off between negativity and dependence on past configurations, i.e. quantum dynamics can be represented either by Markovian quasi-stochastic maps or by positive stochastic processes with higher Markov order. We discuss how such representations of quantum dynamics can be interpreted as stochastic walks through a memory space in the spirit of Projective Simulation, a model of learning and agency in which decisions arise from random walks over an episodic memory network. We further outline how finite-order stochastic kernels can approximate such quantum deliberation processes and show in what regimes the classical machine learning model is recovered.  
\end{abstract}

\maketitle

\tableofcontents

\newpage
\section{Introduction}
Quantum machine learning (QML) models \cite{Biamonte2017QML} define families of input–output maps through parametrised quantum evolution followed by measurement. For instance, a common approach consists of parametrised quantum circuits (PQCs) trained using variational algorithms. \cite{Schuld2018PQC, Havlicek2019Kernel}
While such models can exhibit computational advantages, their internal functioning resists a classical probabilistic interpretation. In classical Markovian stochastic models, processes can be decomposed into sequences of transitions between well-defined configurations, and the resulting probabilities satisfy the Chapman–Kolmogorov divisibility condition (see, e.g. \cite{GardinerStochastic}). That is,  the probability of an outcome can be expressed as a sum over probabilities of intermediate configurations.

Quantum models generically violate that condition. As a consequence, quantum machine learning models lack an interpretation in terms of stochastic deliberation or trajectories. This raises the question of whether quantum processes can nevertheless be represented as stochastic dynamics on an appropriately chosen state space, and, if so, what features of the representation encode the non-classical structure of the underlying model. 

In this work, we address this question by systematically constructing stochastic representations of quantum dynamics based on different choices of configuration spaces. Concretely, we consider representations in which the state of the system is mapped to a probability distribution over a set of configurations defined by a fixed collection of positive operators, and general quantum channels induce linear transition maps on these distributions.

We show that the properties of the resulting stochastic process depend crucially on the choice of representation. Informationally complete representations lead to first-order (Markovian) dynamics, but generally require transition kernels with negative entries. The resulting evolution remains well-defined on the space of configuration probabilities, even though the intermediate transition structure is no longer strictly stochastic. Conversely, representations based on projective measurements preserve positivity of transition probabilities and hence admit an interpretation in terms of stochastic trajectories, but generically require higher-order (non-Markovian) processes and therefore exhibit Chapman–Kolmogorov indivisibility. This establishes a trade-off between positivity and divisibility: quantum dynamics can be represented either through Markovian dynamics with quasi-stochastic transition kernels, or through positive stochastic processes with non-trivial dependence on past configurations.

Moreover, we apply ideas underlying the framework of Projective Simulation (PS) to the stochastic representations of quantum evolution in order to study their interpretability. PS is a reinforcement learning model whose dynamics is formulated in terms of stochastic random walks on a network of memory states. In its classical formulation, the model admits a direct interpretation in terms of stochastic trajectories, thereby providing a transparent account of the agent’s deliberation process. Standard quantum versions generally no longer admit such a trajectory-based description, since the induced probabilities fail to satisfy Chapman–Kolmogorov divisibility. Restricting to the projective representation, we show that these quantum dynamics can nevertheless be represented by positive stochastic processes which are generically non-Markovian, thereby recovering a notion of trajectories at the level of a generalized form of deliberation histories.

We further consider to what extent the resulting stochastic descriptions in PS can approximate quantum dynamics by processes that depend on past configurations with a \textit{fixed finite horizon}. Rather than allowing the effective order of the process to grow with the evolution time, we focus on representations in which this order is bounded. In particular, we discuss how periodicity properties of the underlying dynamics can be used to construct such finite-order approximations. This suggests that, at least in relevant regimes, quantum processes admit stochastic descriptions that retain positivity while requiring only a limited extended dependence on past configurations.

The paper is structured as follows. In Sec. \ref{sec:QML}, we begin by outlining quantum machine learning models and the loss of stochastic interpretability due to the failure of Chapman–Kolmogorov divisibility at the level of intermediate configurations. In Sec. \ref{sec:probabilistic_representations}, we introduce a general framework for probabilistic representations of quantum dynamics based on fixed sets of positive operators, and we study how different choices of configuration spaces affect the induced stochastic processes. In particular, we show that informationally complete representations lead to Markovian but quasi-stochastic dynamics, whereas projective representations yield positive but generally non-Markovian processes, establishing a trade-off between negativity and dependence on past configurations. We illustrate in Sec. \ref{sec:examples} the different stochastic representations in a simple qubit example and also apply the projective representation to stabilizer quantum mechanics. In Sec. \ref{sec:PS}, we apply the projective representation to Projective Simulation, interpreting both classical and quantum variants within a unified stochastic framework and introducing corresponding non-Markovian extensions. There, we also discuss classical limits and potential approximation schemes. We conclude with a discussion of the conceptual implications of the framework, including its relation to quasi-probability representations, stochastic interpretations of quantum dynamics, and the possibility of finite-horizon stochastic descriptions of quantum evolution (Sec. \ref{sec:discussion}).

\section{Quantum Machine Learning and Loss of Stochastic Interpretability}
\label{sec:QML}

In its most common form, QML models consists of a sequence of parameter-dependent unitary operations acting on an initial state, followed by a measurement that produces classical output statistics. Given input data $x$, an initial quantum state $\rho_0$ and parameters $\vec{\theta}$, a parametrized quantum circuit defines a probability distribution over outcomes $c$ of the form
\begin{equation}
\Pr(c \mid x, \vec{\theta}) = \Tr(E_cU(x,\vec{\theta}) \rho_0 U^\dagger(x,\vec{\theta})),
\end{equation}
where $U(x,\vec{\theta})$ is a unitary circuit and $\{E_c\}_c$ is a POVM describing the measurement.

From a probabilistic perspective, this defines a stochastic input–output model from preparations $\rho_0$ to distributions over outcomes $c$.  However, unlike Markovian processes, the internal structure of the computation does not admit an interpretation in terms of probabilistic transitions through \textit{intermediate} states. The Chapman--Kolmogorov condition
\begin{equation}
\Pr(c_{n+1} \mid c_1)
=
\sum_{c_n}
\Pr(c_{n+1} \mid c_n)\Pr(c_n \mid c_1), \label{eq:CK_condition}
\end{equation}
is thus generally not satisfied by the resulting distributions where the $c_n$ define some suitable outcome configurations of the process associated to the POVM $\{E_c\}_c$ time-indexed by $n$. 

For example, for a sequence of unitary evolutions $U^n$ and a projective POVM $\{\ket{c}\bra{c}\}_c$, the probability of observing configuration $c_{n+1}$ given an initial configuration $c_1$,
\begin{equation}
\Pr(c_{n+1} \mid c_1) = \left| \bra{c_{n+1}} U^n \ket{c_1} \right|^2,
\end{equation}
cannot, in general, be expanded as in \cref{eq:CK_condition}.

This violation of the Chapman--Kolmogorov condition reflects the fact that quantum evolution does not define a classical divisible process over intermediate configurations.

As a consequence, quantum machine learning models lack an interpretation in terms of stochastic trajectories or deliberation paths through a configuration space. While the input–output behaviour is probabilistic, the internal computation cannot be decomposed into a sequence of probabilistically meaningful intermediate steps. This poses a fundamental challenge for interpretability, i.e.\ there is, in general, no underlying classical divisible process whose paths can be identified with the evolution of the model.

This observation motivates the central question of this work: Can quantum processes nevertheless be represented as stochastic dynamics on a suitably chosen configuration space, and if so, which features of such representations capture the non-classical features responsible for the quantum nature of such probabilistic descriptions?

\section{Probabilistic Representations of Quantum Machine Learning Models}
\label{sec:probabilistic_representations}

The starting point is a stochastic representation of dynamics on a finite configuration space. Depending on the chosen representation, the induced evolution may either take the form of a first-order linear map on probability distributions or require higher-order stochastic kernels with explicit dependence on a longer sequence of past configurations.


To make this precise, we associate the internal configurations of the model with quantum measurement operators. The most general way to do this is to chose a positive operator-valued measure (POVM), i.e. positive operator frames that are normalised. This guarantees that the configuration variables define a bona fide probability distribution. No measurement is actually performed; the POVM is used only to represent the probabilistic state through its associated Born probabilities. This guarantees that the induced dynamics acts linearly on the probability simplex. As a result, probabilities of a given process will remain positive and normalised, but no restriction is posed on the transition kernels. 

It turns out that any spanning POVM induces a linear Markovian (but generally quasi-stochastic) representation of quantum channels. Non-spanning POVMs, e.g. such as PVMs induce effective non-Markovian stochastic processes.

This yields the salient distinction in the representations of the quantum processes. Informational completeness leads to closure, i.e. the current configuration probabilities contain enough information to determine the next configuration probabilities under any quantum channel. The price is that the dual frame need not be positive, so the resulting transition kernel is generally quasi-stochastic. By contrast, incomplete positive representations, such as projective measurements in a fixed basis, preserve a direct stochastic interpretation of the configuration probabilities but do not generally close under quantum dynamics. The missing information reappears as non-Markovianity. Both of these features distinguish quantum processes from classical processes (see Table \ref{tab:trade-off} for an overview). 
\begin{table}[h]
    \centering
\begin{tabular}{c|c|c}
\textbf{Config. space} & \textbf{Positivity} & \textbf{Markov order} \\
\hline
IC-POVM & No (quasi-stochastic) & $L=1$ \\
Projectors & Yes (stochastic) & $L>1$ (indivisible)
\end{tabular}
\caption{Comparison of stochastic representations induced by different operator choices: informationally complete POVMs yield Markovian but quasi-stochastic dynamics, whereas projective representations yield positive but generally non-Markovian processes.}
\label{tab:trade-off}
\end{table}

We now construct the transition kernels from quantum evolution.  let $\{E_i\}_{i=1}^N$ be a fixed POVM associated with the finite configuration space $V=\{1, ..., N\}$. For a quantum state $\rho$, define the corresponding probability distribution $\Pr(c_k=i) := \Tr(\rho E_i)$, where $c_k\in V$ denotes a configuration at discrete time step $k$.

Let $\mathcal{E}$ be a quantum channel (completely positive trace-preserving map). Then the induced evolution on the configuration probabilities is linear 
\begin{equation}
\label{eqn:general_transition_process}
    \Pr(c_{n+1}=i)=\Tr\!\big(E_i\mathcal{E}(\rho_n) \big)=\sum_{j=1}^N T^{i}_{j}(\mathcal{E})\, \Pr(c_n=j),
\end{equation} where the transition kernel is defined by
\begin{align}
    T^{i}_{j}(\mathcal{E}):=\Tr\!\big(E_i\mathcal{E}(F_j)\big),
\end{align}
where $\{F_j\}_j$ are dual operators to $\{E_i\}_i$.\footnote{Note that, in general, the POVM need not form an operator basis. Let $\mathsf{S}=\operatorname{span}\{E_i\}_{i=1}^N$ and let $\{F_j\}_{j=1}^N$ be a dual frame on $\mathsf{S}$, i.e., $X=\sum_j \Tr(XE_j)F_j$ for every $X\in\mathsf{S}$. Hence, every admissible state $\rho\in\mathsf{S}$ admits the reconstruction $\rho=\sum_j p_jF_j$, where $p_j=\Tr(\rho E_j)$.}
The kernel $T(\mathcal{E})$ satisfies the normalization condition $\sum_{i} T^{i}_{j}(\mathcal{E}) = 1, \forall j,$ but in general $T^{i}_{j}(\mathcal{E})$ is not positive so that $T(\mathcal{E})$ is in general a ``quasi-stochastic'' rather than stochastic transition kernel. However, although the transition kernel $T$ features negativity, it is still the case that all probabilities over configurations remain positive. This follows since the sum in \cref{eqn:general_transition_process} is not an arbitrary linear combination, but a trace of a positive operator. To see this note that $T$ is applied to probability distributions $p$ from a  \textit{subset} $\Omega$
of the probability simplex where $p\in\Omega$ if and only if there exists a state $\rho$ such that $p=(p_i)_i$ with $p_i=\Tr (\rho E_i)$. That is, the transition kernels do not produce negative probabilities because the kernel is only applied to probability vectors that correspond to valid quantum states, for which the linear update is equivalent to a positive quantum evolution. In fact, we have by definition that $T(\Omega)\subset\Omega$.

The representation above shows that quantum dynamics can be expressed as a linear evolution on a space of probabilities once a POVM $\{E_i\}$ is fixed. The properties of the induced transition kernel $T(\mathcal{E})$ depend crucially on this choice. We distinguish three natural regimes in the following.

\subsection{General IC-POVMs representations}
\label{sec:IC_POVM_representation}
We now turn to informationally complete representations. Let $\{E_i\}_{i=1}^N$ be an informationally complete POVM. Since the POVM is informationally complete, for any state $\rho_n$, there exists a dual frame $\{F_j\}_{j=1}^N$ such that $\rho_n = \sum_{j=1}^N \Pr(c_n=j)\, F_j.$

Then the induced transition kernel
\begin{equation}
T^{i}_{j}(\mathcal{E})
=
\Tr\!\big(E_i\mathcal{E}(F_j)\big)
\end{equation}
satisfies $\sum_i T^{i}_{j}(\mathcal{E}) = 1,$ but in general $T^{i}_{j}(\mathcal{E}) \not\ge 0.$

The induced evolution on configuration probabilities permit Markovian processes since for any two quantum channels $\mathcal{E}_1$ and $\mathcal{E}_2$, we have that
\begin{equation}
T^{i}_{j}(\mathcal{E}_2\circ \mathcal{E}_1)
=
\sum_k
T^{i}_{k}(\mathcal{E}_2)
T^{k}_{j}(\mathcal{E}_1).
\end{equation}
To see this, observe that by informational completeness, every operator $X$ that acts on the Hilbert space of the system admits the expansion
\begin{equation}
X=\sum_k \Tr(XE_k)F_k.
\end{equation}
Apply this to $X=\mathcal{E}_1(F_j)$, i.e. $\mathcal{E}_1(F_j)= \sum_k\Tr\!\big(E_k\mathcal{E}_1(F_j)\big)F_k = \sum_k T^{k}_{j} (\mathcal{E}_1)F_k.$ Then, applying $\mathcal{E}_2$ gives $(\mathcal{E}_2\circ\mathcal{E}_1)(F_j)=\sum_kT^{k}_{j}(\mathcal{E}_1)\mathcal{E}_2(F_k).$ Multiplying by $E_i$ and taking the trace yields $T^{i}_{j}(\mathcal{E}_2\circ\mathcal{E}_1)
= \Tr\!\big(E_i(\mathcal{E}_2\circ\mathcal{E}_1)(F_j)\big) = \sum_k T^{k}_{j}(\mathcal{E}_1) \Tr\!\big(E_i\mathcal{E}_2(F_k)\big) = \sum_k
T^{i}_{k}(\mathcal{E}_2)T^{k}_{j}(\mathcal{E}_1).$

The IC-POVM representation thus satisfies the Chapman--Kolmogorov composition law and the dynamics has \emph{Markovian} realizers but only \emph{quasi-stochastic}. The non-classicality of the quantum channel can be interpreted as being encoded in the negativity of the kernel.\footnote{Note that certain restricted sub-theories of quantum mechanics allow for \textit{positive} representations such as the Wigner representation of Gaussian quantum mechanics \cite{Ferrie2011, Hudson1974}, cf.~also the reconstruction from classical Liouville mechanics subject to an epistemic restriction \cite{BartlettRudolphSpekkens2012}.} 

\subsection{SIC-POVMs representation}
For symmetric informationally complete POVMs (SIC-POVMs\footnote{A SIC-POVM in dimension $d$ consists of $d^2$ operators $E_i=\Pi_i/d$, where $\Pi_i$ are rank-one projectors satisfying  $\sum_i E_i=\mathds{1}$ and $\Tr(\Pi_i\Pi_j)=1/(d+1)$ for all $i\neq j$.}), the structure of the kernel becomes particularly transparent. Writing $N=d^2$ and using the canonical dual frame $F_i=d(d+1)E_i-\mathds{1}$, the transition kernel takes the form
\begin{equation}
\label{eqn:SIC_POVM_representation}
T^{i}_{j}(\mathcal{E})
=
d(d+1)\Tr\!\big(E_i\mathcal{E}(E_j)\big) - \frac{1}{d}.
\end{equation}

Again,
\begin{equation}
\sum_i T^{i}_{j}(\mathcal{E}) = 1,\nonumber
\end{equation}
but negativity is generically present. Thus, SIC-POVMs yield a particularly
simple \emph{Markovian} but \emph{quasi-stochastic} representation of quantum dynamics.

In this representation, the quantum channel is encoded as a linear map on the
probability simplex, with deviations from classical stochasticity captured by negative entries of $T$. Among informationally complete representations, SIC-POVMs yield a symmetric and canonical quasi-stochastic representation of quantum dynamics. In this case, the induced transition kernel takes the form of a positive overlap term combined with a uniform negative offset. In particular, in the case of unitary evolutions the first term in \cref{eqn:SIC_POVM_representation} can be understood as a rescaled doubly-stochastic kernel (see, e.g. \cite{Fuchs_quantum_Bayesian_coherence}).\footnote{SIC-POVMs have played a central role in QBist approaches to quantum theory. In that framework, quantum states are interpreted directly as probability distributions over the outcomes of a fixed SIC measurement, and the Born rule takes the form of a modified law of total probability relating different measurement contexts \cite{Fuchs2010QBism,Fuchs_quantum_Bayesian_coherence}.}

In contrast, generic IC-POVMs exhibit negativity that depends on arbitrary choices of dual frames. Moreover, the corresponding transition kernel does not generally reduce to a rescaled stochastic matrix plus a uniform offset as in \cref{eqn:SIC_POVM_representation}.


\subsection{Projective representation and non-Markovianity}

\label{sec:projective_representation}

Projective representations for quantum dynamics arises if we constrain the configurations to projectors. That is, let $\{P_i\}$ be a set of orthogonal projectors $P_i = |i\rangle\langle i|$.
Then one obtains the configuration representation with
\begin{equation}
\Pr(c_n=i) = \Tr(P_i\rho_n),
\end{equation} and induced kernel
\begin{equation}
T^{i}_{j}(\mathcal{E})
=
\Tr\!\big(P_i\mathcal{E}(P_j)\big).
\end{equation}

In this case, the transition kernel can be chosen to be \emph{strictly
stochastic},
\begin{equation}
T^{i}_{j} \ge 0,
\qquad
\sum_i T^{i}_{j} = 1.
\end{equation}

Although the projective POVM is not informationally complete, the representation still reproduces the relevant Born probabilities for the chosen configurations. However, in general it cannot be written as a first-order (Markovian) process. 
Instead, one requires higher-order kernels
\begin{equation}
T^{i}_{j_1,\dots,j_L}
=
\Pr(c_{n+1}=i \mid c_n=j_1,\dots,c_{n-L+1}=j_L),
\end{equation}
with $L>1$, reflecting the failure of Chapman--Kolmogorov divisibility. We give a simple example of such a kernel for the dynamics of a qubit in Section \ref{sec:examples}.

It is important to distinguish divisibility of the induced probability dynamics from Markovianity of a particular stochastic realization. Divisibility merely requires that the evolution of probability distributions factorizes into intermediate transition maps, and therefore guarantees the existence of at least one first-order Markov realizer. However, the same divisible dynamics may in principle also admit non-Markovian realizers. By contrast, if the dynamics is \emph{indivisible}, no Markov realizer exists (cf.\ also \cite{barandes2025quantumsystemsindivisiblestochastic}).

It is also worthwhile to mention that quantum dynamics is naturally formulated in continuous time for which there is no canonical notion of Markov order. Therefore, the Markov order of a stochastic representation of quantum dynamics is not an intrinsic property of the underlying physical process alone, but depends on the chosen temporal coarse-graining. In this work, we explicitly adopt a discrete-time description, and all statements about Markovianity and memory depth are understood relative to this choice.
Thus, projective representations yield \emph{positive but generally non-Markovian}
(indivisible) stochastic processes. More concretely, in general the kernel can be defined to be of the form
\begin{equation}
T^{c_n}_{c_{n-1}, \dots, c_1} =T^{c_{n}}_{c_1}:=\Tr(P_{c_n}\mathcal{E}(P_{c_1})),
\end{equation} i.e. the Markov order is maximal---reaching back to the initial configuration. But there is no dependency on intermediate steps. 

The projective representation becomes Markovian precisely for those channels $\mathcal{E}$ for which diagonal elements evolve independently of off-diagonal elements in the projective basis. For two channels $\mathcal{E}_1$ and $\mathcal{E}_2$, the induced process is divisible if 
\begin{equation}
T(\mathcal{E}_2 \circ \mathcal{E}_1)
=
T(\mathcal{E}_2)\, T(\mathcal{E}_1).
\end{equation}

The left-hand side is
\begin{equation}
\label{eqn:projective_markov_case}
T^{i}_{j}(\mathcal{E}_2 \circ \mathcal{E}_1)
=
\Tr\!\big(P_i\mathcal{E}_2(\mathcal{E}_1(P_j))\big),
\end{equation}
while the right-hand side reads
\begin{equation}
\sum_k
T^{i}_{k}(\mathcal{E}_2)\,
T^{k}_{j}(\mathcal{E}_1)
=
\sum_k
\Tr\!\big(P_i\mathcal{E}_2(P_k)\big)\,
\Tr\!\big(P_k\mathcal{E}_1(P_j)\big).
\end{equation}

Introducing the diagonal (classical) part of $\mathcal{E}_1(P_j)$, i.e. $\rho_{\mathrm{diag}}:= \sum_k \Tr\big(P_k\mathcal{E}_1(P_j)\big)\, P_k$, we see that divisibility holds if and only if
\begin{equation}
\Tr\big(P_i\mathcal{E}_2(\mathcal{E}_1(P_j))\big)
=
\Tr\big(P_i\mathcal{E}_2(\rho_{\mathrm{diag}})\big)
\end{equation}
for all $i,j$.

Equivalently, divisibility holds when the probabilities in the projective basis evolve independently of the off-diagonal components of the input state. In that case the dynamics closes on the projective probability vector, the Chapman--Kolmogorov condition is satisfied, and the induced process admits a Markovian stochastic representation.

This condition admits a natural physical interpretation. The recovery of a Markovian stochastic process in the projective representation requires that the dynamics does not generate coherences between the configuration states, or that such coherences are effectively suppressed at each step. This is also the situation realised when the system is coupled to an environment that induces decoherence and dephasing in the configuration basis. In this case the channel preserves diagonal states in the configuration basis, i.e. it maps classical probability distributions to classical probability distributions such that  the evolution can be understood as a sequence of classical transitions between configurations.

The projective representation therefore yields positive but generally non-Markovian stochastic processes. A particularly important instance of this structure is provided by the stochastic–quantum correspondence, to which we turn now.

\paragraph{Projective representations and stochastic-quantum correspondence.}

Recent work by Barandes \cite{Barandes_2025} proposes a \textit{stochastic--quantum correspondence}, according to which a quantum system can be understood as a stochastic process unfolding in an ordinary configuration space governed by ``indivisible'' (i.e.\ non-Markovian) stochastic laws, with Hilbert space structures playing only a secondary, representational role. Within this perspective, the case of the so-called \textit{unistochastic} construction can be understood as a special case of the present projective representation. According to the stochastic--quantum correspondence, the configuration space is given by a fixed orthonormal basis, i.e.\ a projective decomposition $\{P_i\}$ and a unitary quantum evolution. The resulting stochastic kernel is then defined directly in terms of transition probabilities $\Pr(c_n \mid c_1) = |\langle c_n | U^n | c_1 \rangle|^2$, yielding a strictly positive but generally non-Markovian process whose Markov order is effectively unbound and scales with the temporal extent of the evolution. Within our framework, this corresponds precisely to choosing a projective space of configurations and a channel for which the induced transition maps are positive but fail to satisfy Chapman--Kolmogorov divisibility. Positivity is preserved at the expense of divisibility. 

In contrast, as we showed, more general choices of operator bases (e.g.\ informationally complete POVMs) or channels allow one to interpolate between positivity and Markovianity, making explicit that the indivisible stochastic description is not a separate mechanism, but a particular representation within a broader class of probabilistic models of quantum dynamics. It should be noted, however, that although it describes non-Markovian processes, the projective representation is a fully probabilistic dynamics that preserves positivity and thus an interpretation in terms of stochastic processes.  

\section{Examples}
\label{sec:examples}
\subsection{Stochastic representations of qubit dynamics}
\label{sec:qubit_example}

We now illustrate the projective and informationally complete stochastic representation for the simplest non-trivial configuration space. 

Let $V=\{0,1\}$ be a finite configuration space consisting of two configurations $c\in V$, associated with the orthonormal basis states $\ket{0},\ket{1}\in\mathbb{C}^2$. The projective representation is therefore defined with respect to the projectors $E_c=\ket{c}\!\bra{c}, c\in V.$

Consider a unitary evolution generated by the Hamiltonian $H=\frac{\hbar\omega}{2}\sigma_x$, which generates a coherent oscillation between the two configurations. Here, $\omega$ is the angular frequency setting the rate of the oscillation, and $\sigma_x$
 is the Pauli-$x$ matrix. Let $U_{\tau}$ denote the unitary evolution over a total time interval $\tau$, and introduce a discrete-time evolution step by defining
\begin{equation}
U(\Delta t):=\sqrt[N]{U_{\tau}}:=e^{-\frac{i}{\hbar}H\Delta t},
\qquad
\Delta t=\frac{\tau}{N},
\end{equation}
such that $U_{\tau} = U(\Delta t)^N$.

In the configuration basis, the discrete-time unitary takes the form
\begin{equation}
U(\Delta t)
=
\begin{pmatrix}
\cos\left(\frac{\omega\Delta t}{2}\right)
&
-i\sin\left(\frac{\omega\Delta t}{2}\right)
\\
-i\sin\left(\frac{\omega\Delta t}{2}\right)
&
\cos\left(\frac{\omega\Delta t}{2}\right)
\end{pmatrix},
\end{equation} which entails the transition kernel
\begin{equation}
T^{c_n}_{c_1}
=
|\langle c_n|U(\Delta t)^n|c_1\rangle|^2.
\end{equation}

But this kernel does not satisfy the Chapman--Kolmogorov condition.\footnote{This is, of course, well known from the basic phenomenon of, e.g., single particle quantum interference.}  Indeed, a first-order stochastic process would require
\begin{equation}
|U(2\Delta t)_{10}|^2
=
\sum_{k=0,1}
|U(\Delta t)_{1k}|^2
|U(\Delta t)_{k0}|^2,
\end{equation}
which is clearly not satisfied. The unitary evolution therefore cannot be represented by a positive first-order stochastic process on the configuration space. The resulting process is thus non-Markovian as the transition probabilities depend on the initial configuration of the process rather than only on the most recent configuration.

For comparison, consider an informationally complete representation of the same qubit dynamics using a SIC-POVM $\{E_c\}_{c\in V}$, where the configuration space now consists of four outcomes $V=\{1,2,3,4\}$. The corresponding configuration probabilities are  $\Pr(c_n=c):=\Tr(\rho_n E_c)$. Writing the quantum state in Bloch form as
$\rho_n=\frac{1}{2}\bigl(\mathds{1}+\mathbf{r}_n\cdot\boldsymbol{\sigma}\bigr)$,
the SIC probabilities determine the Bloch vector uniquely. The unitary evolution $
\rho_{n+1}=U(\Delta t)\rho_nU^\dagger(\Delta t)$ therefore induces a rotation $\mathbf{r}_{n+1}=R\,\mathbf{r}_n$ on the Bloch sphere. It can be shown that $\Pr(c_{n+1}=i)= \sum_{j\in V} T^{i}_{j}\,
\Pr(c_n=j)$, with
\begin{equation}
T^{i}_{j}
=
\frac{1}{4}
\Bigl(
1+3\,\mathbf r_i\!\cdot R\mathbf r_j
\Bigr),
\end{equation}
where $\mathbf r_i$ denote the tetrahedral Bloch vectors defining the SIC-POVM. The resulting dynamics is therefore first-order and satisfies the divisibility condition. In contrast to the projective representation, no history dependence is required because the probability distribution $\Pr(c_n)$ already contains a complete description of the quantum state. However, the entries of $T$ are generally not positive, so that the induced dynamics is quasi-stochastic rather than genuinely stochastic.
This does not lead to negative configuration probabilities, because the physically admissible SIC distributions occupy only a proper subset of that simplex, and the quasi-stochastic evolution maps this subset into itself.

\subsection{Projective representation of stabilizer dynamics}

Stabilizer quantum mechanics provides a useful special case in which the Markov order required by a positive projective representation is finite and small. Consider a single qudit of prime dimension $d$. We describe the system in the discrete phase-space representation, where a quantum state $\rho$ is represented by a discrete Wigner function $W_\rho(q,p)$ on $\mathbb Z_d^2$. The coordinates $q,p\in\mathbb Z_d$ play the role of finite position and momentum coordinates, and label phase-space points $(q,p)\in\mathbb Z_d^2$. Let the projective basis be the position basis, so that the configuration is $c := q$.

Clifford unitaries act on the discrete phase space by affine symplectic transformations. Ignoring affine shifts for simplicity, the evolution is 
\begin{equation}
\begin{pmatrix}
q_{n+1} \\ p_{n+1}
\end{pmatrix}
=
S
\begin{pmatrix}
q_n \\ p_n
\end{pmatrix},
\qquad
S =
\begin{pmatrix}
A & B \\
C & D
\end{pmatrix}
\in \mathrm{Sp}(2,\mathbb{Z}_d).
\end{equation}
where $\mathrm{Sp}(2,\mathbb{Z}_d)$ is the
symplectic group over $\mathbb{Z}_d$, i.e., the group of matrices preserving the discrete position-momentum relations.
Although the full phase-space dynamics is first-order Markovian, the projective representation only retains the position coordinate $q_n$ and discards the hidden momentum $p_n$. The resulting visible dynamics is therefore not generally first-order Markovian.

However, since $S$ is a $2\times 2$ symplectic matrix, it satisfies its characteristic equation $S^2 - \Tr(S)S + \mathds{1}=0 \pmod d$, with $\det(S)=1 \pmod d$. Applying this to the phase-space trajectory implies that the coordinate satisfies $q_{n+1}=\Tr(S)q_n - q_{n-1} \pmod d$. Thus, the effective stochastic process on the projective representation closes at Markov order $L=2$. It is represented by the deterministic stochastic kernel \begin{equation}
T^{i}_{j_2,j_1}
=
\delta_{i,\, \Tr(S)j_2-j_1 \!\!\!\pmod d}.
\end{equation}
Hence, for single-qudit stabilizer dynamics, the projective representation is positive and non-Markovian, but only of finite Markov order $2$.

For $N$ qudits, the phase space has dimension $2N$. A similar argument gives a finite-order recurrence for the position coordinates, but the generic Markov order is bounded by $2N$ rather than $2$.\footnote{For \(N\) qudits, write the phase-space trajectory as
\(x_n=(\mathbf q_n,\mathbf p_n)^{T}\in\mathbb Z_d^{2N}\), with \(x_{n+1}=Sx_n\) and \(S\in\operatorname{Sp}(2N,\mathbb Z_d)\).
If \(\chi_S(\lambda)=\lambda^{2N}+a_1\lambda^{2N-1} +\cdots+a_{2N}\), the Cayley--Hamilton theorem gives \(S^{2N}+\sum_{\ell=1}^{2N}a_\ell S^{2N-\ell}=0\). Applying this to the trajectory and projecting onto the position coordinates yields $\mathbf q_{n+1}
=-\sum_{\ell=1}^{2N}a_\ell\mathbf q_{n+1-\ell} \pmod d.$ Hence the next configuration is determined by at most the preceding \(2N\) configurations, so the induced process admits a realization of Markov order \(L\leq 2N\). This bound need not be tight. Writing \(S=\left(\begin{smallmatrix}A&B\\ C&D\end{smallmatrix}\right)\), if \(B\) is invertible over \(\mathbb Z_d\), then \(\mathbf p_n=B^{-1}(\mathbf q_{n+1}-A\mathbf q_n)\). Thus two consecutive configurations determine the full phase-space state, and the projected dynamics closes already at order \(L\leq2\).}
\section{Interpreting Quantum Models via Generalized Projective Simulation}
\label{sec:PS}
Following the general discussion of probabilistic representations of quantum dynamics, we now show how such representations can be used to analyze quantum dynamics in a concrete quantum learning model. To this end, we consider a stochastic model of deliberation and learning, namely projective simulation (PS) \cite{Briegel2012ProjectiveSimulation}, where deliberative processes are represented by random walks on an adaptive memory network. In its standard (classical) formulation, the network encodes an agent's memory structure, deliberation corresponds to stochastic propagation through this network, and learning is implemented through updates of the associated transition weights. The probabilistic representations introduced above can then be viewed as generalized projective-simulation models, where quantum dynamics determines the effective propagation rules (for deliberation) and learning corresponds to their modification over time.



In PS, the agent's memory is represented by a directed graph $G=(V,E)$, whose vertices $V$ are referred to as clips, reflecting their interpretation as elementary fragments of episodic memory. The percepts $S\subset V$ are the input nodes activated by environmental stimuli. The actions $A\subset V$ are output nodes through which the agent acts on its environment. The edges $E$ carry weights $h_{ij}$, which represent the strength of association between internal clips $c_i$ and clip $c_j$. The probability of transitioning from clip $c_j$ to $c_i$ is given by the stochastic matrix $P_{ij} = \frac{h_{ij}}{\sum_{j'} h_{ij'}}.$

Deliberation begins with the excitation of a percept clip. The excitation then propagates through the clip network as a time-homogeneous discrete-time random walk with transition matrix $P_{ij}$, until an action clip is reached.

This random-walk process induces a statistical input-output map from percepts to actions, called the policy function. For a percept $s\in S$ and an action $a \in A$, the policy function $\Pr(a\mid s)$ of standard PS is then given by \cite{briegel2025projective}
\begin{align}
\Pr(a \mid s)
&= \sum_{n=1}^{\infty} \sum_{\pi_G(n)}
  O\!\left(a \mid c_{k_n}\right)
  \prod_{j=0}^{n-1}
    \Pr\!\left(c_{k_{j+1}} \mid c_{k_j}\right)
  I\!\left(c_{k_0} \mid s\right) .
\label{eq:ps-classical}
\end{align}
Here, $n$ denotes the length of the deliberation path, and $\pi_G(n)$ denotes the set of all length-$n$ paths
$\pi=(c_{k_0},c_{k_1},\ldots,c_{k_n})$
on the clip network $G$, with $(c_{k_j},c_{k_{j+1}})\in E$ for all $j=0,\ldots,n-1$. Here, $j$ indexes the position along a path, whereas $k_j$ denotes the label of the clip occupied at that position.
The factor $I(c_{k_0}\mid s)$ describes the interface from the environment to the agent's  memory, specifying the probability that percept $s$ initializes the internal random walk at clip $c_{k_0}$.
The product gives the probability of following the path $\pi$ through the clip network under the internal transition probabilities $\Pr(c_{k_{j+1}}\mid c_{k_j})$.
Finally, $O(a\mid c_{k_n})$ describes the interface from the agent's  memory back to the environment, specifying the probability that the terminal clip $c_{k_n}$ gives rise to action $a$.
The policy $\Pr(a\mid s)$ is therefore obtained by summing over all internal deliberation paths that connect the percept $s$ to the action $a$.

The interpretability of PS has been demonstrated in several settings, ranging from classical learning agents with explicitly traceable clip-network dynamics \cite{Briegel2012ProjectiveSimulation,mautner2015projective,pazem2025free}, to concept formation and transparent artificial intelligence \cite{eva2023minimal}, and to quantum-information applications in which learned policies correspond to explicit adaptive controls, protocols, or experimental arrangements \cite{tiersch2015adaptive,wallnofer2020machine,melnikov2018active}. Interpretability is achieved insofar as the policy is not represented only as an input-output function, but is mediated by an explicit clip network whose nodes and transition weights can be inspected. The semantic content of clips is inherited from their construction and functional role within the memory: percept and action clips are tied to the agent's interface with the environment, while intermediate clips represent memory fragments---typically remembered percepts or actions or compositions thereof. Random-walk trajectories through this network can therefore be interpreted as deliberation paths, and learning can be analyzed through changes in the transition structure.

To use PS as an interpretative framework for quantum models, we relax the standard stochastic and time-homogeneous Markov dynamics in several directions. We emphasize that the following construction is not intended as a unique quantum generalization of the PS framework, but should rather be understood as one possible extension.

First, informationally complete probabilistic representations of quantum dynamics may induce quasi-stochastic maps, which are not positive on the full simplex but preserve a restricted admissible subset (cf. the IC-POVM representation in \Cref{sec:IC_POVM_representation}).\footnote{Note that quasi-stochastic refers to the linear transition maps, i.e. the non-positivity of transition probabilities and not the probabilities themselves. The probabilistic states are always positive (cf. also the discussion on quasi-probability representations in \Cref{sec:discussion}.} Secondly, projective or otherwise incomplete representations may lead to effective non-Markovian dynamics, where the next state depends on a finite history rather than only on the current state (cf. the projective representation in \Cref{sec:projective_representation}). Thirdly, QML models are typically composed of sequences of different quantum maps, for instance gates, measurements, or parameter updates, making a time-inhomogeneous description more natural than a single fixed transition rule. We therefore introduce two corresponding generalizations of the standard PS deliberation dynamics. An overview of the different representations of the PS framework are found in \Cref{tab:ps-path-conditions}.

\begin{table}[t]
\centering
\footnotesize
\begin{tabular}{@{}lllll@{}}
\textbf{Representation} & \textbf{Markovian} & \textbf{Positivity} &  \textbf{Interpretation}  & \\
\hline
Classical PS & yes & yes & clip random walk \\
IC/SIC-POVM & yes & no & path in $K\subsetneq\Delta_V$ \\
Projective & no & yes & clip random walk \\
\hline
\end{tabular}
\caption{Two conditions for path interpretability in generalized PS.}
\label{tab:ps-path-conditions}
\end{table}

More concretely, we first allow the probabilistic state space of the clip network to be restricted to a  closed convex subset $K\subseteq \Delta_V$ of the full probability simplex over clips. Standard PS is recovered in the special case $K=\Delta_V$, where the transition matrix $P$ is stochastic and satisfies $P(\Delta_V)\subseteq \Delta_V$. For a restricted admissible set $K\subsetneq\Delta_V$, we instead allow more general linear maps $T$ that preserve only the admissible region,
\begin{align}
T(K)\subseteq K ,
\end{align}
without requiring $T(\Delta_V)\subseteq\Delta_V$. Thus, $T$ may fail to be stochastic and to constitute positive transition probabilities on the simplex, while still defining a valid evolution on the admissible PS states. For restricted admissible sets $K$ it helps to think of them as configurations themselves.
For instance, this is the case for the IC-POVM representation in Section \ref{sec:IC_POVM_representation}.


Second, we allow for indivisible (and hence non-Markovian) deliberation dynamics. In this case, the next admissible distribution need not depend only on the current distribution $p_n\in K$, but may depend on a finite history
$(p_n,p_{n-1},\ldots,p_{n-\ell+1})$. It is useful to regard the relevant state as the history vector
\begin{align}
\eta_n
=
(p_n,p_{n-1},\ldots,p_{n-\ell+1})\in K^\ell .
\end{align}
The evolution is then described by an admissibility-preserving history map
\begin{align}
T^{(\ell)}:K^\ell \longrightarrow K,
\qquad
p_{n+1}
=
T^{(\ell)}(\eta_n),
\end{align}
with $T^{(\ell)}(\eta_n)\in K$ for all $\eta_n\in K^\ell$. The Markovian case is recovered for $\ell=1$.

Third, we allow for time-inhomogeneous dynamics. The update rule may depend explicitly on the deliberation step $n$, so that a family of admissibility-preserving maps is used instead of a single time-independent map:
\begin{align}
T_n(K)\subseteq K,
\qquad
p_{n+1}=T_n p_n .
\end{align}
More generally, in the non-Markovian case one may allow a time-dependent history map
\begin{align}
T^{(\ell)}_n:K^\ell\longrightarrow K,
\qquad
p_{n+1}
=
T^{(\ell)}_n(p_n,p_{n-1},\ldots,p_{n-\ell+1}) .
\end{align}
This generalization is particularly natural for QML models, where the effective dynamics is often given by a sequence of distinct operations rather than by repeated application of a single map. In this case, the relevant propagation rule changes with the deliberation step: each map $T_n$ describes the effective transition associated with a particular layer or measurement step.

Standard stochastic PS is recovered by taking $K=\Delta_V$, $\ell=1$, and $T_n=T$ independent of $n$, with $T$ stochastic; in this case $T$ is identified with the usual PS transition matrix $P$.

With the generalization in place, we can now ask in which sense the resulting dynamics permits the path-based interpretation in terms of random clip walks that is central to standard PS. It is useful to decompose the states in $K$ into extremal points. That is, one may write
\begin{align}
p_j = \sum_{\lambda} \mu_j(\lambda)\, e_\lambda,
\qquad e_\lambda\in \operatorname{Ext}(K),
\end{align}
or, more generally, use an integral over $\operatorname{Ext}(K)$ when the set of extremal points is continuous.

In standard PS, as well as in the projective representation, one has effectively $K=\Delta_V$. Hence, every admissible state has the unique decomposition 
\begin{equation}
    p=\sum_{c\in V}p(c)\delta_c,
\end{equation} where the extremal points $\delta_c$ are definite clip states. If the dynamics induces positive transition probabilities between these extremal states, the process can be interpreted as a random walk over clips. Standard PS is the Markovian realization of this, i.e. $l=1$. The projective quantum representation preserves the interpretation in terms of random walks over clips, but the transition probabilities generally depend on the preceeding history, i.e. $l>1$ in general.

The stochastic process over extremal points then reduces uniquely to a random walk over clips,
\begin{align}
\delta_{c_{k_0}} \longrightarrow \delta_{c_{k_1}} \longrightarrow \cdots \longrightarrow \delta_{c_{k_n}},
\end{align} recovering the standard clip-path interpretation of PS.

In informationally complete representations, by contrast, the admissible set $K=\{p_i=\Tr(\rho E_i)\}$ is a proper subset of the ambient simplex $\Delta_V$. The simplex vertices $\delta_c$ are generally not elements of $K$, and the linear map on the these coordinates is not positive on those vertices. Thus, the POVM labels should not be interpreted as definite clips being activated during deliberation along a path.

Instead, the elements $p\in K$ themselves can be viewed as the states of the generalized process. The induced dynamics is then a Markovian transition on the state space $K$ (in the deterministic case), i.e.
\begin{align}
p_0 \longrightarrow p_1 \longrightarrow \cdots \longrightarrow p_n,
\qquad p_j\in K\subseteq \Delta_V , \label{eq:simplex_path}
\end{align} where each $p_j(c)$ specifies the weight assigned to clip $c$ at step $j$.\footnote{More generally, stochastic dynamics on the admissible state space $K$ one has a probability measure $\mu_n$ over admissible probability vectors $p\in K$, and the dynamics is specified by a Markov kernel $M_n(dp'|p)$ on $K$ such that $\mu_{n+1}(B)=\int_K M_n(B|p)\,\mu_n(dp), B\subseteq K.$ The deterministic IC-POVM evolution $p_{n+1}=T_n p_n$ is recovered as the special case $M_n(dp'|p)=\delta_{T_n p}(dp')$. The effective clip-probability vector associated with $\mu_n$ is $\bar p_n=\int_K p\,\mu_n(dp).$ Thus, the stochasticity is now a probability distribution over admissible probabilistic states, rather than a transition probability between definite clips.} But this is a trajectory of admissible probability vectors rather than a random walk over elementary clips. Unless \(K\) is a simplex, the decompositions into extremal points in $\operatorname{Ext}(K)$ are generally non-unique. The resulting extremal-path description therefore is not fixed by the stochastic representation itself.

Thus, in IC-POVM representations the word ``clip'' should be understood as a coordinate label or reference outcome, not as an elementary memory state that can be sharply occupied. The representation is tomographically complete and Markovian at the level of admissible probability states, but it does not by itself supply a canonical random walk over definite clips. This is still sufficient for a network-based description of how percepts are transformed into action probabilities. However, it is not in general a trajectory through definite clip excitations. Unlike standard PS paths through definite clip excitations, \cref{eq:simplex_path} describes a trajectory of distributed states in $K\subseteq\Delta_V$.

In sum, in classical/projective PS we have stochastic processes over definite clip states, whereas in informationally complete PS the stochastic process is over admissible probability states.





For time-homogeneous stochastic PS, deliberation is Markovian: the current clip, or more generally the current distribution over clips, is a sufficient state variable for predicting the next step. This yields the Chapman--Kolmogorov divisibility condition
\begin{align}
(P^{n+m})_{ji}
=
\sum_{k\in V}
(P^m)_{jk}(P^n)_{ki},
\end{align}
and supports a local form of step-by-step explainability: each transition can be interpreted in terms of the presently occupied clip and its outgoing transition weights. 

The non-Markovian generalization changes the relevant explanatory unit. If the dynamics has memory length $\ell$, the next admissible state is determined not by $p_n$ alone, but by an augmented state
\begin{align}
\eta_n
=
(p_n,p_{n-1},\ldots,p_{n-\ell+1}),
\end{align}
so that
\begin{align}
p_{n+1}
=
T^{(\ell)}(\eta_n).
\end{align}
The process can therefore be regarded as Markovian on the enlarged history space, but non-Markovian after projection back to the clip-distribution space. Interpretability is correspondingly shifted from local clip-to-clip transitions to history-dependent transition rules: an explanation of the next step must specify not only the current excitation pattern, but also the relevant deliberation context encoded in the preceding trajectory. 

Regarding the input and output coupling, in standard PS, the percept and action interfaces are implemented by distinguished subsets of clips. A percept $s$ initializes the deliberation process by exciting a percept clip, while reaching an action clip $a\in A\subset V$ terminates the random walk and produces the corresponding output. In the generalized informationally complete setting, where the admissible states form a restricted convex set $K\subseteq\Delta_V$, sharply localized clip excitations $\delta_c$ are not admissible. The input and output interfaces must therefore be lifted from clips to admissible distributions. The in-coupling is described by a map
\begin{align}
I:S\to K,
\qquad
s\mapsto p_0=I(\cdot|s),
\end{align}
which assigns to each percept an admissible initial distribution over the clip network. 

In the standard setting of PS, the deliberation process is connected to the action space by an out-coupling function $O(a\mid c)$, which assigns action probabilities to terminal clips. In the generalized setting, where the internal deliberation state is an admissible distribution $p\in K\subseteq\Delta_V$ rather than a definite clip, the natural analogue is a readout map
\begin{align}
O(a\mid p):K\to[0,1],
\qquad
\sum_{a\in A} O(a\mid p)=1 .
\end{align}
This map assigns action probabilities directly to admissible distributed clip states.  A generalized policy can therefore be obtained by summing over admissible deliberation trajectories and applying $O(a\mid p_n)$ to the final distribution $p_n$.

Taken together, these generalizations define an enlarged projective-simulation framework with two levels of interpretation. Informationally complete representations provide a distribution-level dynamics on admissible probability states, while projective representations retain a genuine stochastic path interpretation, generally at the cost of history dependence. Standard PS is recovered when \(K=\Delta_V\), the dynamics is stochastic, Markovian, and time-homogeneous, and the extremal admissible states are the delta distributions over clips. The generalized framework retains the central interpretative structure of PS---namely, a network-based account of how inputs are transformed into action probabilities---while allowing to accommodate quasi-stochastic representations and history-dependent dynamics. It therefore provides a common language in which classical PS, probabilistic representations of quantum dynamics, and QML models composed of sequences of quantum maps can be compared at the level of their induced deliberation dynamics.

\section{Discussion}
\label{sec:discussion}
The main result of this work is that the stochastic representation of quantum dynamics exhibits a trade-off between positivity and divisibility. Informationally complete representations allow one to express quantum evolution as a first-order stochastic process on a space of probabilities, preserving Chapman--Kolmogorov divisibility and hence Markovianity. However, the corresponding transition kernels are generically not positive and therefore define quasi-stochastic rather than genuinely stochastic dynamics. By contrast, projective representations preserve positivity of transition probabilities but generically violate divisibility, requiring stochastic processes with non-trivial history dependence. In this case, quantum dynamics can only be represented by higher-order stochastic processes whose transition structure depends on extended histories.

From this perspective, negativity and non-Markovianity appear as two complementary ways of encoding quantum interference within a stochastic description. Informationally complete representations retain divisibility by allowing negative transition weights, whereas projective representations retain positivity by shifting the non-classical structure into temporal correlations. The framework developed here therefore suggests that the obstruction to a fully classical stochastic description of quantum dynamics is not tied uniquely either to negativity or to non-Markovianity alone, but rather to the impossibility of simultaneously maintaining positivity and first-order divisibility. 

\paragraph{Stochastic interpretations of quantum learning models.}
A central motivation for the present framework is the question of interpretability in quantum machine learning models. Classical learning models can be interpreted in terms of stochastic state transitions that admit an explicit trajectory-based picture. By contrast, quantum machine learning models typically evolve through coherent processes that do not admit a straightforward interpretation in terms of stochastic trajectories on a configuration space. In particular, the induced probabilities generally fail to satisfy the Chapman--Kolmogorov condition, obstructing an interpretation in terms of divisible stochastic processes.

The projective representation developed in this work provides a way of recovering a stochastic description of such dynamics while retaining positivity of probabilities. Applied to quantum Projective Simulation, this yields a stochastic interpretation of quantum deliberation processes in terms of extended histories rather than single-step trajectories. Classical PS already provides an interpretable reinforcement learning model formulated through stochastic random walks on a clip network. Within the projective representation, the resulting quantum dynamics can be understood as positive but generally non-Markovian stochastic processes. In this sense, the framework developed here provides a possible route toward interpreting certain classes of quantum machine learning dynamics in terms of stochastic processes of this kind. 

\paragraph{Relation to other representations.} The stochastic representations discussed in this work are closely related to quasi-probability approaches to quantum theory. In informationally complete representations, quantum channels induce linear evolutions on probability vectors through kernels that are generally not positive. The resulting dynamics is therefore represented by first-order quasi-stochastic processes, where the non-classicality of the evolution appears through negative transition weights rather than through negative observable probabilities. Importantly, however, the propagated distributions remain genuine (that is, non-negative) probability distributions over configurations, in contrast both to approaches that explicitly allow negative probabilities over underlying configurations \cite{al2013simulating,kaszlikowski2021little,morris2022witnessing,onggadinata2026ideal} and to quasi-probability representations such as discrete Wigner functions, where negativity is commonly regarded as a signature of non-classicality \cite{Hudson1974, BartlettRudolphSpekkens2012, ChiribellaSpekkens2016}. Indeed, the physically realised probability distributions remain positive since the kernels act only on the restricted subset of the probability simplex corresponding to valid quantum states.

Moreover, our perspective is closely related to the framework of generalized probabilistic theories, whose modern operational form was developed in quantum foundations and quantum information as a common language for classical, quantum, and post-quantum probabilistic theories in terms of generalized probabilistic states (see, e.g. \cite{Hardy2001,Barrett2007}).

The stochastic representations considered in this work should not be confused with hidden-variable theories. Although one may formally interpret the configuration space $V$ as a space of underlying configurations, the stochastic representations introduced here do not specify the physical realization of the corresponding dynamics. In the context of Projective Simulation, for example, they represent macroscopic memory states (clips) of an agent.

Consequently, properties such as non-locality, contextuality, or dynamical stability cannot in general be inferred from the stochastic representation itself. A given stochastic process may admit multiple physical realizations with different causal structures. The present framework should therefore be understood as a representation of quantum probability dynamics rather than as a commitment to a particular underlying ontology. Additional physical assumptions are required before conclusions about locality or the microscopic structure of the dynamics can be drawn \cite{FankhauserMyrvoldIndivisibleRelativity}.

\paragraph{Relation to the ``stochastic--quantum correspondence''.} The projective stochastic representations considered in this work contain the recent stochastic--quantum correspondence developed by Barandes as a special case \cite{Barandes_2025}. In that approach, quantum systems are described as positive stochastic processes on configuration space that generically violate divisibility. There, the standard Markovian stochastic evolution is replaced with non-Markovian stochastic laws. Within the framework developed here, this corresponds to \textit{projective} representations. 

The broader perspective developed in the present work shows that such indivisible stochastic representations arise as one particular realization within a larger family of stochastic representations of quantum dynamics. In particular, the comparison with informationally complete representations demonstrates that one may alternatively preserve first-order divisibility at the expense of positivity of the transition structure. Nevertheless, the projective stochastic representations underlying the stochastic--quantum correspondence are particularly interesting because the positivity of the induced stochastic process is not preserved under a straightforward generalization from projective measurements to arbitrary POVMs. A related generalization of the stochastic---quantum correspondence was recently put forward by Doukas \cite{doukas2026emergencequantummechanicsstochastic}. 

\paragraph{Finite-history approximations of quantum dynamics.} An interesting open question is to what extent positive stochastic representations of quantum dynamics can be compressed to finite-order processes over finite time horizons.

The exact projective representations discussed above generally require history dependence that grows with the temporal extent of the process. It is therefore natural to ask whether positive stochastic representations of quantum dynamics can nevertheless admit effective finite-history descriptions in physically relevant regimes.

A suggestive observation comes from the structure of finite-dimensional unitary dynamics. Let $U_T$ be a unitary evolution over a fixed time interval $T$, and define a discrete-time evolution by introducing a time step $\Delta t = T/N$ together with a unitary root $U(\Delta t):=\sqrt[N]{U_T}$,
such that $U_T = U(\Delta t)^N.$ The corresponding configuration-space probabilities $q_\nu(n)=|\langle \nu|U(\Delta t)^n|c_0\rangle|^2$
then form periodic sequences determined by the eigenphase differences of $U(\Delta t)$. Diagonalizing $U(\Delta t)=\sum_r e^{i\theta_r}\Pi_r$, one obtains $q_\nu(n)=\sum_{r,s} A^{(\nu)}_{rs} e^{in(\theta_r-\theta_s)}$, for suitable coefficients $A^{(\nu)}_{rs}$. The probability sequences are therefore exponential functions of discrete time and satisfy finite recurrence relations when the eigenphases are rational or approximated by rational number.

This observation suggests that the apparent temporal complexity of quantum dynamics may in some cases admit compression into finite-history stochastic descriptions. In particular, when the eigenphase differences are periodic, or approximately periodic over finite horizons, the corresponding probability dynamics can exhibit effective recurrences. Such recurrences naturally suggest stochastic representations with bounded history dependence. More generally, the quasiperiodic structure of finite-dimensional quantum evolution raises the question of whether finite-order stochastic kernels may approximate the induced probability dynamics over finite horizons.

The stabilizer example discussed above provides a concrete illustration of this phenomenon: there the effective stochastic description closes at finite Markov order due to the underlying symplectic structure of the dynamics. More generally, positive phase-space representations such as those appearing in Gaussian and stabilizer quantum mechanics suggest that restricted sectors of quantum theory may admit finite-history positive stochastic realizations.

Understanding when such compressions are possible, and how the required history dependence depends on the structure of the dynamics, the temporal coarse-graining, and the chosen representation, remains an open problem.

Furthermore, an open question is whether quantum computational resources can be characterized in terms of the Markov order required by positive stochastic representations. For example, stabilizer quantum mechanics admits a
finite-order stochastic description, suggesting that restricted quantum resources may correspond to bounded-memory stochastic processes, while universal quantum computation may require arbitrarily large effective history dependence. A related question is whether finite-history stochastic approximations to quantum circuits can themselves be implemented or emulated by quantum circuits with reduced resources with bounded effective Markov order.



\section{Conclusions}
\label{sec:conclusion}

In this work, we investigated how quantum machine learning models can be represented and interpreted in stochastic terms. Starting from fixed configuration spaces associated with positive operator representations, we showed that quantum dynamics induces corresponding probabilistic processes whose properties depend on the chosen representation.

Our analysis confirms that there is no unique stochastic description of a quantum process. Informationally complete representations yield closed first-order dynamics on probability distributions, but generally require quasi-stochastic transition kernels. Projective representations, by contrast, preserve positivity and admit an interpretation in terms of stochastic trajectories, but typically do so only through history-dependent processes. Quantum dynamics therefore does not single out a unique notion of stochasticity; rather, different representations distribute the non-classical features of the theory between negativity and temporal dependence in different ways.

This perspective provides a common framework for comparing classical stochastic models, probabilistic representations of quantum theory, and quantum machine learning models. In particular, it allows quantum versions of Projective Simulation to be analysed in terms of stochastic deliberation processes, extending the trajectory-based intuition underlying the classical model. More generally, the framework suggests that questions of interpretability in quantum learning models may be fruitfully reformulated as questions about the existence and properties of stochastic representations.

Several open directions remain. An important question is how to characterize the minimal history dependence required by positive stochastic representations of quantum dynamics and to understand its dependency on the underlying dynamics, temporal coarse-graining, and approximation accuracy. More broadly, the framework developed here may suggest a new way of analysing quantum resources through stochastic processes that represent them, potentially providing a bridge between quantum dynamics, learning models, and the theory of non-Markovian stochastic systems.


\acknowledgements 
This research was funded in whole or in part by the Austrian Science Fund (FWF), and the Wittgenstein Award [SFB BeyondC F7102, DOI:10.55776/F71; WIT9503323, DOI: 10.55776/WIT9503323]. For open access purposes, the authors have applied a CC BY public copyright license to any author-accepted manuscript version arising from this submission. We gratefully acknowledge support from the European Union (ERC Advanced Grant, QuantAI, No. 101055129). The views and opinions expressed in this article are however those of the author(s) only and do not necessarily reflect those of the European Union or the European Research Council - neither the European Union nor the granting authority can be held responsible for them. We thank Jacob Barandes and Philip A. LeMaitre for helpful discussion on a draft of this work. 

\bibliography{library}
\appendix

\end{document}